\documentclass[journal=mamobx,manuscript=article,layout=twocolumn]{achemso}

\usepackage[font=footnotesize,labelfont=bf,labelsep=period]{caption}   
\usepackage[font=footnotesize,labelfont=bf,labelsep=period]{subcaption}                     

\usepackage[version=3]{mhchem} 
\usepackage[T1]{fontenc}       
\usepackage{graphicx} 
\usepackage{epstopdf} 
\usepackage{color}
\usepackage[colorinlistoftodos]{todonotes}
\PassOptionsToPackage{english}{babel}
\usepackage{amssymb,amsfonts,amsmath,cleveref}
\usepackage{newtxmath}
\usepackage{bbold}
\usepackage[normalem]{ulem}
\usepackage{float}
\usepackage{morefloats}
\usepackage{soul}
\usepackage[english]{babel}
\usepackage{multirow}
\usepackage{placeins}
\usepackage{makecell}

\newcommand{\revision}{\textcolor{black} }
\newcommand{\revDY}{\textcolor{black} }
\newcommand{\revIP}{\textcolor{black} }
\SectionNumbersOn

\author{Sriteja Mantha}
\affiliation{Division of Chemistry and Chemical Engineering, California Institute of Technology, Pasadena, California 91125, USA}
\altaffiliation{Contributed equally to this work}

\author{Alec Glisman}
\affiliation{Division of Chemistry and Chemical Engineering, California Institute of Technology, Pasadena, California 91125,USA}
\altaffiliation{Contributed equally to this work}

\author{Decai Yu}
\affiliation{The Dow Chemical Company, Core R\&D, 633 Washington St., Midland, Michigan 48674, USA}

\author{Eric Wasserman}
\affiliation{The Dow Chemical Company, Consumer Solutions R\&D, 400 Arcola Road, Collegeville, Pennsylvania 19426, USA}

\author{Scott Backer}
\affiliation{The Dow Chemical Company, Consumer Solutions R\&D, 400 Arcola Road, Collegeville, Pennsylvania 19426, USA}

\author{Zhen-Gang Wang}
\email{zgw@caltech.edu}
\affiliation{Division of Chemistry and Chemical Engineering, California Institute of Technology, Pasadena, California 91125, USA}

\title[An \textsf{achemso} demo]
  {Adsorption isotherm and mechanism of \ce{Ca^{2+}} binding to polyelectrolyte}
\keywords{Adsorption isotherm, Antiscalant polyelectrolytes, Enhanced sampling molecular simulations}
\begin{document}


\begin{abstract}
 
\revDY{Polyelectrolytes, such as polyacrylic acid (PAA)}, \revision{can} effectively mitigate \ce{CaCO3} scale formation.
Despite their success as antiscalants, the underlying mechanism of \ce{Ca^{2+}} binding to polyelectrolyte chains remains unresolved. Through all-atom molecular dynamics simulations, we construct an adsorption isotherm of \ce{Ca^2+} binding to sodium polyacrylate (\ce{NaPAA}) and investigate the associated binding mechanism. We find that the number of calcium ions adsorbed [\ce{Ca^{2+}_{ads}}] to the polymer saturates at moderately high concentrations of free calcium ions [\ce{Ca^{2+}_{aq}}] in the solution. This saturation value is intricately \revision{connected} with the binding modes accessible to \ce{Ca^{2+}} ions when they bind to the polyelectrolyte chain. We identify two dominant binding modes:
\revision{the first involves binding to at most two carboxylate oxygens on a polyacrylate chain, and the second, termed the high binding mode, involves binding to four or more carboxylate oxygens. As the concentration of free calcium ions [\ce{Ca^{2+}_{aq}}] increases from low to moderate levels, the polyelectrolyte chain undergoes a conformational transition from an extended coil to a hairpin-like structure, enhancing the accessibility to the high binding mode. At moderate concentrations of [\ce{Ca^{2+}_{aq}}], the high binding mode accounts for at least a third of all binding events. The chain's conformational change and its consequent access to the high binding mode is found to increase the overall \ce{Ca^{2+}} ion binding capacity of the polyelectrolyte chain.}

\end{abstract}

\section{Introduction}
Divalent \revision{metal} ions, such as \ce{Ca^{2+}}, exhibit a pronounced affinity for dissolved anions like carbonate (\ce{CO3^{2-}}) in an aqueous solution. 
\revDY{The ion pairs, which readily associate, precipitate out of solution and form solid deposits (scale) due to their low solubility limit. Scale formation} presents challenges to residential and industrial piping systems, restricting the fluid flow and fouling components.\cite{Antony2011,Tijing2010,Muryanto2014} These metal ions \revision{also} form complexes with household products, such as detergents, and disrupt their \revision{efficacy}.

\revDY{Polyelectrolytes, such as polyacrylic acid (PAA)}, are commonly employed to mitigate scale formation.\cite{Huang2007, Sinn2004, Yu2004, Aschauer2010, Backer2021} Although it is not fully understood what makes an effective antiscalant polyelectrolyte, a few mechanisms for scale prevention have been proposed. The polyelectrolytes could prevent nucleation via chelating metal ions from the solution.\cite{Gindele2022,Sinn2004} The chelation reduces the concentration of metal ion and the likelihood of their association with the dissolved anions. Simultaneously, polyelectrolytes could also adsorb onto the surfaces of scale crystals and prevent further growth \revDY{or deposition}.\cite{Reddy2001,Yu2004,Aschauer2010, Sparks2015} 

The solution behavior of polyelectrolytes in divalent salt solutions poses important challenges in designing polyelectrolytes with enhanced antiscalant activity, as the polyelectrolyte\revision{--}ion complex itself can precipitate and lead to further scale deposition. \cite{Schweins2001,Schweins2003, Michaeli1960,Muthukumar2017,Schweins2006,Yethiraj2008,Park2012,Mantha2015,Chialvo2005,Dobrynin2006,Prabhu2005,Duan2022} Cloud point measurements revealed instances of phase separation into a polymer-poor (supernatant) liquid and a polymer-rich liquid with addition of divalent ions.\cite{Sabbagh2000} Boisvert \textit{et al.} performed osmotic pressure measurements and \revision{showed} that the solution behavior of polyelectrolytes in divalent salt solutions is primarily influenced by a proposed site-binding mechanism of divalent cations\cite{Boisvert2002}. The site-binding mechanism facilitates bridging of non-adjacent \revDY{repeat units} by the divalent cations. Through a Fourier Transform Infrared (FTIR) dialysis technique, \cite{Fantinel2004} Fantinel \textit{et al.} identified monodentate, bidentate, and bridging modes when \ce{Ca^{2+}} ions bind to carboxylate groups of a polyacrylate chain. However, the relative importance of each of these binding modes---particularly that of the bridging mode---on the ability of polyelectrolyte to chelate \ce{Ca^{2+}} ion has not been elucidated. Furthermore, the influence of \ce{Ca^{2+}} ion concentration on these binding modes remains elusive.

Mean-field theories have shed light on the solution behavior of polyelectrolytes in divalent salt solutions.\cite{Solis2001,Kundagrami2008,Wittmer1995,delaCruz1995,Ermoshkin2003,Lee2009,Deserno1999} In addition to establishing conditions for the precipitation behaviours, these theoretical models have predicted a large reduction in polymer size with addition of divalent ions,\cite{Liu2003} \revision{ beyond what is expected from electrostatic screening}. The chain collapse was primarily attributed to ion bridging between non-neighboring \revDY{repeat units}. Coarse-grained implicit solvent molecular simulations, employing generic models for polyelectrolyte chains and ions, have confirmed the chain collapse and attributed it to the bridging capability of divalent ions.\cite{Zhou2016,Liu2003,Kos2005} All-atom molecular dynamics simulations have gone a step further by explicitly treating solvent molecules to investigate the molecular principles that govern the binding of divalent cations to the polyelectrolyte chain.\cite{Molnar2004,Yao2018,Sappidi2016,Patel2017,Tribello2009,Chung2012,Bulo2007} These investigations have \revision{each} reported that the \ce{Ca^{{2+}}} ion is strongly coordinated with the polyelectrolyte chain, resulting in highly coiled conformations with a chain rigidity reminiscent of crystal-like structures. Due to the strong \ce{Ca^{2+}}\revision{--}polyelectrolyte interactions and the number of \ce{Ca^{2+}} ions binding to the polyelectrolyte chain, these models hint at the overcharging of the \ce{Ca^{2+}}--polyelectrolyte complex. However, recent potentiometric titration assays\cite{Gindele2022} suggest that only $\frac{1}{3}$ of the binding sites are occupied before a fully charged polyelectrolyte chain reaches its saturated value of \ce{Ca^{2+}}-binding capacity.

The saturation value in the adsorption isotherm, \revDY{which describes the maximum amount of} \ce{Ca^{2+}} \revDY{that can be bound to a polyelectrolyte chain, reflects the chelating capacity} of the polyelectrolyte chain. However, the molecular principles that govern the corresponding adsorption isotherm have not been addressed. Specifically, the \revision{interplay} between the site-binding nature of \ce{Ca^{2+}}--polyelectrolyte interactions and the conformational transitions of the polyelectrolyte, aimed at enhancing both the chelating capacity and the solubility of the \ce{Ca^{2+}}--polyelectrolyte complex, has not been explored.

In this study, we address the mechanism of \ce{Ca^{2+}} adsorption onto a polyacrylate chain. We construct an adsorption isotherm to describe the binding behaviors of \ce{Ca^{2+}} ions to polyacrylate. We then determine the different binding modes accessible for \ce{Ca^{2+}} binding to the chain and quantify their impact on the adsorption isotherm. In our follow-up manuscript,\cite{Glisman2023} we address questions related to \ce{Ca^{2+}} ion-mediated association between polyelectrolyte chains. The rest of the manuscript is organized as follows. In \Cref{Sec:Methods}, we present a Hamiltonian Replica Exchange Molecular Dynamics (HREMD) protocol to selectively bias \ce{Ca^{2+}}-polyelectrolyte interactions and efficiently sample the configurational space. We then introduce a free energy perturbation approach coupled with molecular dynamics to compute the adsorption isotherm describing \ce{Ca^{2+}} binding to the polyelectrolyte chain. We discuss the results of these calculations in \Cref{Sec:Results} and present the conclusions in \Cref{Sec:Concl}.

\section{Models and Methods}
\label{Sec:Methods}
 We investigated the mechanism of calcium ion binding to polyacrylate chain using all-atom molecular dynamics (MD) simulations. Our simulation system consists of a single \revDY{polyacrylic acid} (PAA) chain with 32 \revDY{repeat units}, solvated in a cubic water box with an edge length of 12 nm.  All \revDY{repeat units} on the polymer are charged, consistent with the solution conditions for antiscalant activity(i.e., solution pH $\thicksim 10$). Sodium ions were added for electroneutrality. The average end-to-end distance of such a polymer was $\thicksim 5$ nm. We chose our box dimensions so that the polymer does not interact with its periodic image. In \Cref{Tab:Prm}, we report compositions of different systems studied in this work.

\begin{table}
\centering
\begin{tabular}{|l|c|c|c|c|}\hline
$\mathbf{Label}$ & $\mathbf{Ca^{2+}}$ & $\mathbf{Cl^{-}}$ & $\mathbf{Na^{+}}$ & $\mathbf{Water}$  \\ \hline
\revision{0\ce{CaCl_2}} & 0 & 0 & 32 & 56,448\\ \hline
4\ce{CaCl_2} & 4 & 8 & 32 & 56,436\\ \hline
8\ce{CaCl_2} & 8 & 16 & 32 & 56,424 \\ \hline
16\ce{CaCl_2} & 16 & 32 & 32 & 56,400\\ \hline
32\ce{CaCl_2} & 32 & 64 & 32 & 56,352 \\ \hline
64\ce{CaCl_2} & 64 & 128 & 32 & 56,256\\ \hline
96\ce{CaCl_2} & 96 & 192 & 32 & 56,160 \\ \hline
128\ce{CaCl_2} & 128 & 256 & 32 & 56,064 \\ \hline
\end{tabular}
\caption{Composition of the cubic simulation box with an edge length of 12 nm, containing a single \revision{sodium} polyacrylate chain with 32 \revDY{repeat units}, and at various numbers of \ce{CaCl2} in water. }
\label{Tab:Prm}
\end{table}


\textbf{Force field choice and the importance of solvent electronic polarization:} 
We employed the Generalized AMBER Force Field (GAFF) with the SPC/E water model, which had been rigorously tested by Mintis et al. for modeling the polyacrylate chain. \cite{Mintis2019} During our modeling of calcium ions, we observed that the ``full'' electrostatic charge force field parameters overestimated calcium ion binding to carboxylate groups on the polyacrylate chain, resulting in a charge inversion of the polymer chain inconsistent with experimental reports (see Supporting Information). Dubou\'e-Dijon et al., who investigated calcium ion binding to insulin, reported that such an overestimation was due to inadequate treatment of electronic dielectric screening when using full charges on ions with non-polarizable forcefields.\cite{DubouDijon2018} The correct energetics of ion-pair formation could be captured by molecular models with explicit polarization.\cite{Bedrov2019} However, a general-purpose force field with explicit polarization, tested to reproduce the properties of polyacrylates, was not readily available, and polarizeable force fields introduce large computational expense.

Alternatively, in our models, we included electronic dielectric screening by uniformly scaling the charges of all solute atoms. This approach, known as the electronic continuum charge correction (ECC), is a mean-field method that attempts to mimic charge-carrying species within an electronic dielectric continuum.\cite{DubouDijon2020, Leontyev2011} While the ECC scheme is a physically meaningful concept, it is primarily used as an \textit{ad hoc} solution to incorporate electronic polarization into an otherwise non-polarizable force field. ECC schemes tend to fail in systems with a discontinuity in the high-frequency dielectric constant.\cite{Kirby2019} Moreover, force field parameterizations implicitly account for electronic polarization effects to some extent. Introducing additional ECC correction may overly compensate for electronic polarization effects, resulting in a significant underestimation of cohesive energy density and leading to unphysical solution behavior.\cite{Cui2019} Nevertheless, the ECC scheme yields meaningful observations in common electrolyte systems where the high-frequency dielectric constant remains uniform throughout.

Within the scope of our study, the interactions of interest involve calcium ions and carboxylate groups on the polyacrylate chain. Biomolecular systems with the same carboxylate-calcium ion pair have shown success with ECC schemes. \cite{Timr2018,Martinek2018,Kohagen2014,Kohagen2014a} We particularly employ the ECC scheme reported by Jungwirth and coworkers, who provided parameters for modeling calcium and other ions in an aqueous solution.\cite{Martinek2018} In their work, the authors uniformly scaled the charges on ions by a factor of 0.75. Lennard-Jones parameters describing dispersion interactions of corresponding ions were optimized to reproduce \textit{ab initio} molecular dynamics results for ion-pairing and neutron scattering experiments. Such a parameter set accurately described the properties of calcium ions in an aqueous solution and their association with carboxylate groups on amino acids. We used the same parameter set to model calcium and other ions in our simulation system.

When modeling the electronic polarization effects of the polyacrylate chain in an aqueous solution, we scaled the partial charges of all the atoms on the polyacrylate chain by 0.75 and used Lennard-Jones parameters reported by Mintis \textit{et al.} \cite{Mintis2019} to describe dispersion interactions. Although our approach is a commonly accepted practice,\cite{Kohagen2014a} we emphasize that it is not rigorous. However, different properties of polymers strongly depend on their chain length, and finding the right strategy to optimize their dispersion interaction parameters is not obvious.

Since the ``full-charge'' parameters by Mintis \textit{et al.} \cite{Mintis2019} reproduce the structural properties of the polyacrylate chain in a salt-free solution, we validated the predictions of the ``scaled-charge'' model against the former. Even though we did not re-optimize the Lennard-Jones interaction parameters of the polyelectrolyte chain, the conformational flexibility of a polyelectrolyte chain in a salt-free solution did not change. Additionally, we observed identical distributions for structural properties when compared to the full-charge'' model (see Supporting Information).

\textbf{Simulation methodology:} We used GROMACS 2022.3 \cite{Abraham2015, VanDerSpoel2005, Berendsen1995} patched with PLUMED 2.8.1 \citep{Bonomi2009, Tribello2014, Plumed2019} and employed the following protocol to conduct our molecular dynamics simulations of the systems reported in \Cref{Tab:Prm}. First, we minimized the energy of the system using the steepest descent algorithm until the maximum force on any atom in the system was smaller than 100 kJ mol$^{-1}$ nm$^{-1}$. Next, we equilibrated the system at constant NPT conditions with pressure and temperature set to 1 atm and 300 K, respectively. While fluctuations in the energy and box size minimized within a few nanoseconds of the equilibration run, it took tens of nanoseconds to equilibrate the chain structure. We used the Berendsen barostat\cite{Berendsen1984} with the velocity-rescaling stochastic thermostat during the first 10 ns of the equilibration. For the remainder of the equilibration run, we switched to the Parrinello–Rahman barostat\cite{Parrinello1981} with the Nos\'{e}\revision{--}Hoover chain thermostat for the accurate reproduction of the thermodynamic properties of the system.\cite{Nos1984,Hoover1985} Following the equilibration run, we conducted production MD simulations of these systems in the NVT ensemble using the Nos\'{e}\revision{--}Hoover chain thermostat.

We employed the leap-frog time integration algorithm with a finite time step of 2 fs to integrate \revision{the} equations of motion. Additionally, we utilized the LINCS constraint algorithm to convert all bonds with hydrogen atoms into constraints.\cite{Hess1997} We applied periodic boundary conditions along all three spatial axes and used the particle mesh Ewald (PME) method with a minimum Fourier spacing of 0.12 nm to calculate the long-ranged \revision{electrostatic} interactions.\cite{Darden1993,Essmann1995} We applied a cutoff distance of 1.2 nm for computing van der Waals interactions. \revision{We used the same distance as the real-space cut-off value while computing the PME electrostatics}.

\textbf{Need for enhanced sampling molecular simulations:} \revision{Although the ECC scheme with the non-polarizable force field greatly reduced the PAA{--}\ce{Ca^2+} binding/unbinding relaxation times, regular MD simulations were unable to efficiently sample the polymer conformational space in an aqueous \ce{CaCl2} solution. Even a microsecond long trajectory was not sufficient to sample polymer conformations in any of the systems listed in} \Cref{Tab:Prm} with calcium \revision{numbers} higher than 4 \ce{Ca} ions. We direct readers to the Supporting Information document for the relevant data and discussion.

To address the challenges associated with polymer conformational sampling in an aqueous \ce{CaCl2} solution, we employed Hamiltonian Replica Exchange Molecular Dynamics (HREMD).\cite{Wang2011,Bussi2013} Our HREMD framework is based on the flexible implementation of the REST2 variant,\cite{Wang2011} as previously reported by Bussi.\cite{Bussi2013} We introduced a parameter $\lambda$ to selectively bias the interactions between the polymer and ions, as well as the dihedral potential components of the Hamiltonian. The charges of the ions and the polymer were scaled by a factor of $\sqrt{\lambda}$, while their Lennard\revision{-}Jones interaction parameter ($\epsilon$) was scaled by $\lambda$. Similarly, the polymer dihedral potential was also scaled by $\lambda$. With this scheme for $\lambda$-parameterized Hamiltonian, $\lambda=1$ corresponds to the system of interest with full-scale interactions. We determined that the parameterized Hamiltonian with $\lambda=0.67$ rapidly sampled the polymer conformational space.  Coordinate exchange between neighboring replicas were attempted every 500 steps. We utilized 16 replicas, with $\lambda$ values ranging from 1 to 0.67 (geometrically spaced), to simulate polymer conformations in an aqueous solution containing 8 \ce{CaCl2} or 16 \ce{CaCl2}. For higher Ca$^{2+}$ numbers, we increased the number of replicas to 24. This combination of number of replicas and the $\lambda$-range yielded acceptable exchange probabilities ($\thicksim0.3$) between neighboring replicas. All the relevant results reported in the subsequent sections were obtained by averaging over a 250 ns production HREMD run, conducted at constant volume and a temperature of 300 K.

\textbf{Computing ion adsorption isotherm from molecular dynamics simulations:} The primary objective of the current work is to investigate Ca$^{2+}$ chelation onto a model polyacrylate chain. In an aqueous \ce{CaCl2} solution containing a polyacrylate chain, a dynamic equilibrium exists between calcium ions that are freely dispersed in the solution ($\mathrm{Ca^{2+}_{aq. free}}$) and the calcium ions adsorbed per monomer of the polyacrylate chain ($\mathrm{AA-Ca^{2+}}$). 

\begin{equation}
\mathrm{AA} + \mathrm{Ca^{2+}_{aq.free}}\rightleftharpoons \mathrm{AA-Ca^{2+}}
\label{eq:Equil}
\end{equation}

Here, $\mathrm{AA}$ represents the concentration of \revDY{repeat units} on the polymer chain that are not bound to any calcium ions. An adsorption isotherm, quantifying \Cref{eq:Equil}, describes calcium ion chelating ability of a model polyacrylate chain. We constructed the isotherm by plotting the number of calcium ions adsorbed per monomer ($\mathrm{AA-Ca^{2+}}$) against the concentration of calcium ions that are freely dispersed in the solution ($\mathrm{Ca^{2+}_{aq. free}}$).

Computing $\mathrm{AA-Ca^{2+}}$ from the simulation trajectory is straightforward. We calculated the number of calcium ions that are within 0.7 nm (see SI) of the polymer atoms at each frame. This quantity was then divided by the number of \revDY{repeat units} per chain (i.e. 32 in this case) and ensemble averaged over the trajectory.

We determined $\mathrm{Ca^{2+}_{aq. free}}$ by equating the chemical potential of the Ca$^{2+}$ ions in the system ($\mu^\mathrm{System}_{\ce{CaCl2}}$) with that of a pure CaCl$_2$ aqueous suspension ($\mu^\mathrm{Solution}_{\ce{CaCl2}}$).
This required conducting simulations of two separate sets of systems: one containing a polyacrylate chain to determine $\mu^\mathrm{System}_{\ce{CaCl2}}$, and the other without a polyacrylate chain for $\mu^\mathrm{Solution}_{\ce{CaCl2}}$.
We employed the procedure laid out by Panagiotopoulos and coworkers to determine the chemical potentials via free energy perturbation approach.\cite{Panagiotopoulos2020, Saravi2022, Young2018, Young2019}

In brief, $\mu_{\ce{CaCl2}}$ represents the change in free energy when adding (or removing) a unit of \ce{CaCl2},  This is expressed in \Cref{eq:mu} as the sum of the corresponding ideal gas component ($\mu^{\mathrm{id}}_{\mathrm{CaCl_2}}$) and the residual component ($\mu^{\mathrm{R}}_{\mathrm{CaCl_2}}$).

\begin{align}
&\mu_{\ce{CaCl2}}=\mu^{\mathrm{id}}_{\ce{CaCl2}} + \mu^{\mathrm{R}}_{\ce{CaCl2}} \nonumber \\
&\mu^{\mathrm{id}}_{\ce{CaCl2}} = \mu^0_{\ce{Ca^{2+}}} + 2\mu^0_{\ce{Cl^{-}}} + 3RT\ln{\left(\frac{\sqrt[3]{4}N_{\ce{CaCl2}}k_BT}{{P^0\left<V\right>}}\right)} \nonumber \\
&\mu^{\mathrm{R}}_{\ce{CaCl2}} = \mu^\mathrm{R}_\mathrm{vdw} + \mu^\mathrm{R}_\mathrm{coul}
\label{eq:mu}
\end{align}
Here, $\mu^0_{\ce{Cl^{-}}}$ and $\mu^0_{\ce{Ca^{2+}}}$ are the respective chemical potentials of \ce{Cl^-} and \ce{Ca^{2+}} at the reference pressure of 1 bar. We used the tabulated value of $\mu^0_{\ce{Cl^{-}}}$ from the NIST-JANAF thermochemical tables\cite{NIST}. Mou\v{c}ka et al. estimated the value for $\mu^0_{\ce{Ca^{2+}}}$\cite{Mouka2018}, and we employed the same value in our calculations. $\mathrm{N_{CaCl_2}}$ represents the number of \ce{CaCl2} units in the simulation box, $k_B$ denotes the Boltzmann constant, $P^0=1,\mathrm{bar}$ is the reference pressure, and $\left<V\right>$ stands for the average volume obtained from NPT simulations of a system with a specific composition.

To calculate $\mu^{\mathrm{R}}_{\mathrm{CaCl_2}}$, we gradually decoupled  a \ce{Ca^{2+}} and two \ce{Cl^-} from the system. The initial configuration for this study came from the most probable polymer conformation identified from the previously described enhanced sampling simulations. We then tagged a \ce{Ca^{2+}} ion that was adsorbed onto the polymer backbone. In a solution without the polyacrylate chain, a randomly selected \ce{Ca^{2+}} ion served as the tagged ion, and a comparison point between the two systems. Regardless of the system, two randomly chosen \ce{Cl^{-}} ions were tagged.

The electrostatic interactions of the tagged \ce{Ca^{2+}} ion with other particles in the system were gradually turned off in 11 stages ($\phi=1,0.9,\ldots,0$). Here, $\phi=1$ represents the system with fully activated electrostatic interactions of the tagged calcium ion, while $\phi=0$ corresponds to complete deactivation. Each stage consisted of a series of molecular dynamics simulation steps, beginning with energy minimization, followed by 5 ns of equilibration and a 10 ns production simulation run at \revDY{ a pressure of 1 bar and a temperature of 300 K.} The output from the production step of the $i$th stage ($\phi_i$) served as the initial configuration for the MD simulation steps of the $(i+1)$th stage ($\phi_{i+1}$). The same methodology was then applied to deactivate the electrostatic interactions of each tagged chloride ion and also the van der Waals interactions of the tagged calcium ion and the two tagged chloride ions. We then used the Bennett Acceptance Ratio (BAR) \cite{Bennett1976,Mester2015} approach, natively implemented in the GROMACS MD package\cite{Abraham2015}, to estimate the residual chemical potential ($\mu^{\mathrm{R}}_{\ce{CaCl2}}$) from these different stages of MD simulations. \revision{ Notably, given the position-dependent nature of the unbinding free energy of a} \ce{Ca^{2+}} \revision{ion from the polymer backbone, we calculated the unbinding free energy for each adsorbed} \ce{Ca^{2+}} \revision{ion individually and used their average to estimate} $\mu^{\mathrm{R}}_{\ce{CaCl2}}$. 

\section{Results and Discussion}
\label{Sec:Results}
\begin{figure}[htbp]
    \centering
    \includegraphics[width=3 in]{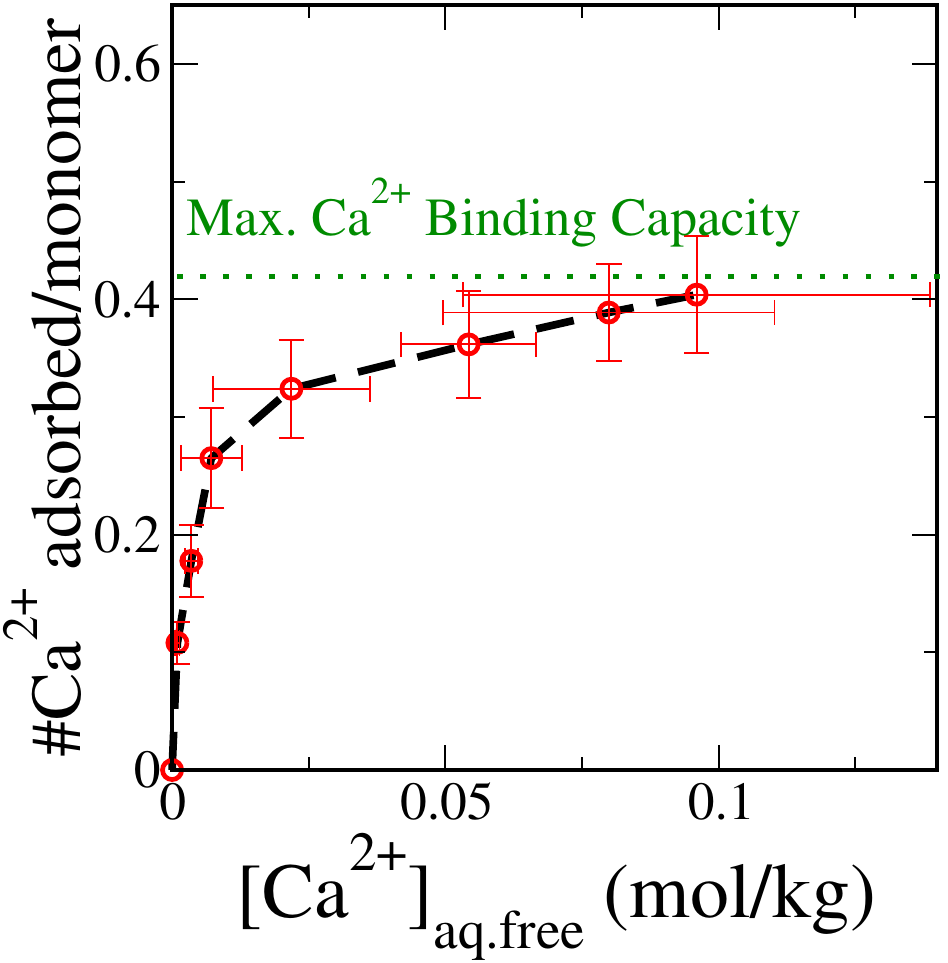}
    \caption{Isotherm describing \ce{Ca^{2+}} ion binding to a polyacrylate chain with 32 \revDY{repeat units}. Error bars represent one standard deviation around the sample mean.\revision{\  Note: The apparent positive slope of the last two data points is a consequence of large uncertainties in determining corresponding solution concentration of free} \ce{Ca^{2+}} \revision{in the system. The differences in their vertical axis values are very minimal.} }
    \label{fig:AdsIso}
\end{figure}

The ability of a polyacrylate chain to sequester \ce{Ca^{2+}} ions is intricately tied to the bulk solution concentration of \ce{Ca^{2+}}, denoted as $\left[\mathrm{Ca^{2+}_{{aq.free}}}\right]$. In \Cref{fig:AdsIso}, we present the adsorption isotherm and describe the extent of \ce{Ca^{2+}} binding onto a 32-mer polyacrylate chain as a function of $\left[\mathrm{Ca^{2+}_{{aq.free}}}\right]$. At low \ce{Ca^{2+}} concentrations, most of the binding sites on the polyacrylate chain are available, and an increase in $\left[\mathrm{Ca^{2+}_{{aq.free}}}\right]$ results in \revision{a rapid} increase in the number of \ce{Ca^{2+}} ions sequestered by the polyacrylate chain. However, at moderate concentrations, the accessible binding sites become occupied, and the polyacrylate chain saturates with \ce{Ca^{2+}}.

From the plateau in the adsorption isotherm, we note that the maximum binding capacity of a model polyacrylate chain is $0.40 \pm 0.05$ \ce{Ca^{2+}} ions per monomer, which aligns with recent potentiometric titration experiments\cite{Gindele2022}. Although the adsorption isotherm bears some resemblance to a Langmuir model, we observe that the \ce{Ca^{2+}} binding does not obey the model assumptions. Notably, the conformational flexibility of the polyelectrolyte chain results in distinct chemical environments around each binding site, rendering the binding sites non-equivalent. 

\begin{figure*}[htbp]
    \centering
    \begin{subfigure}{0.45\linewidth}
        \centering
        \includegraphics[width=\linewidth]{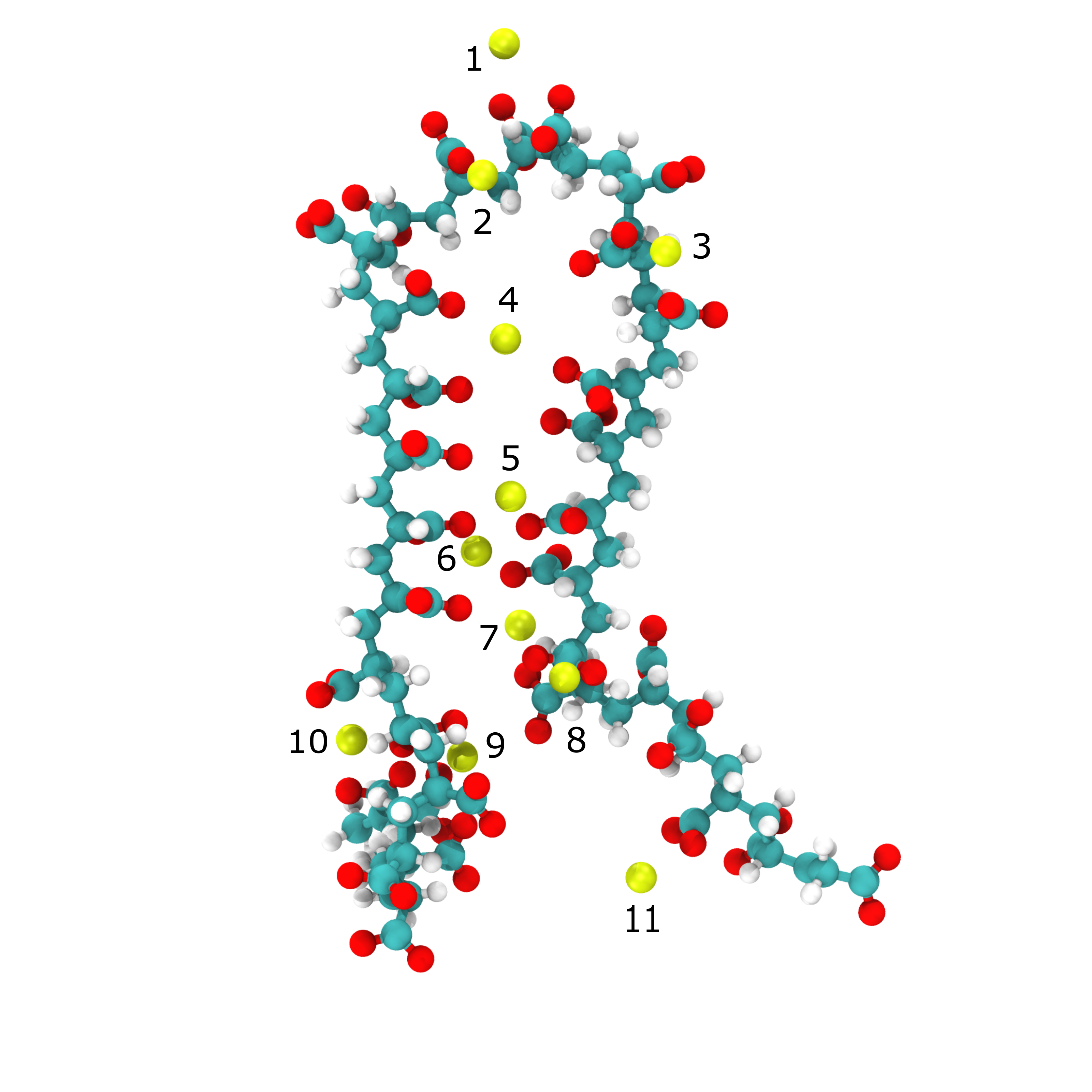}
        \caption{}\label{sfig:BindSt}
    \end{subfigure}
    \hspace{-2eM}
    \begin{subfigure}{0.40\linewidth}
        \centering
        \begin{tabular}{|c|c|}
            \hline
            Label & $\mathrm{\Delta F \left[kJ/mol\right]}$ \\
            \hline
            1 & $814\pm 1$ \\
            2 & $815\pm 1$ \\
            3 & $814\pm 1$ \\
            4 & $818\pm 2$ \\
            5 & $818\pm 2$ \\
            6 & $821\pm 3$ \\
            7 & $821\pm 3$ \\
            8 & $815\pm 1$ \\
            9 & $828\pm 2$ \\
            10 & $823\pm 1$ \\
            11 & $817\pm 1$ \\
            \hline
            \multirow{1}{*}{Average} & \multirow{1}{*}{$818\pm 5$} \\
            \hline
         \end{tabular}
        \caption{}\label{sfig:UnEnr}
    \end{subfigure}
    \caption{
    (\subref{sfig:BindSt}) Conformation of a 32-mer polyacrylate chain with the most probable radius of gyration in a solution corresponding to $\mathrm{\left[Ca^{2+}\right]_{aq.free}=0.026 ~ mol/kg}$
    and (\subref{sfig:UnEnr})
    free energies of unbinding a \ce{Ca^{2+}} ion that is adsorbed to the polyacrylate chain.
    }
    \label{fig:BindDat}
\end{figure*}

We demonstrate this phenomenon in a polyelectrolyte solution corresponding to $\mathrm{Ca^{2+}_{aq.free}=0.026 mol/kg}$. First, we identify the system configuration that corresponds to the polymer chain with the most probable radius of gyration (\Cref{sfig:BindSt}). Then, we independently unbind each of the bound \ce{Ca^{2+}} ions. We employ the free energy perturbation approach described in \Cref{Sec:Methods} to compute the unbinding free energy, and report the corresponding values in \Cref{sfig:UnEnr}.

The unbinding free-energies of the 11 \ce{Ca^{2+}} ions broadly fall into two classes.
We categorize these binding sites into a ``low'' binding mode (${\le 818}$ kJ/mol) and ``high'' binding mode (${> 818}$ kJ/mol). From the binding sites depicted in \Cref{sfig:BindSt} along with the accompanying free energies in \Cref{sfig:UnEnr}, we identify that the high binding mode is approximately 5--10 kJ/mol energetically more favorable than the low binding mode and facilitates a \ce{Ca^{2+}} ion-bridge between non-neighboring carboxylate groups on the polyacrylate chain. 

\begin{figure}[htbp]
    \centering
    \includegraphics[width=3 in]{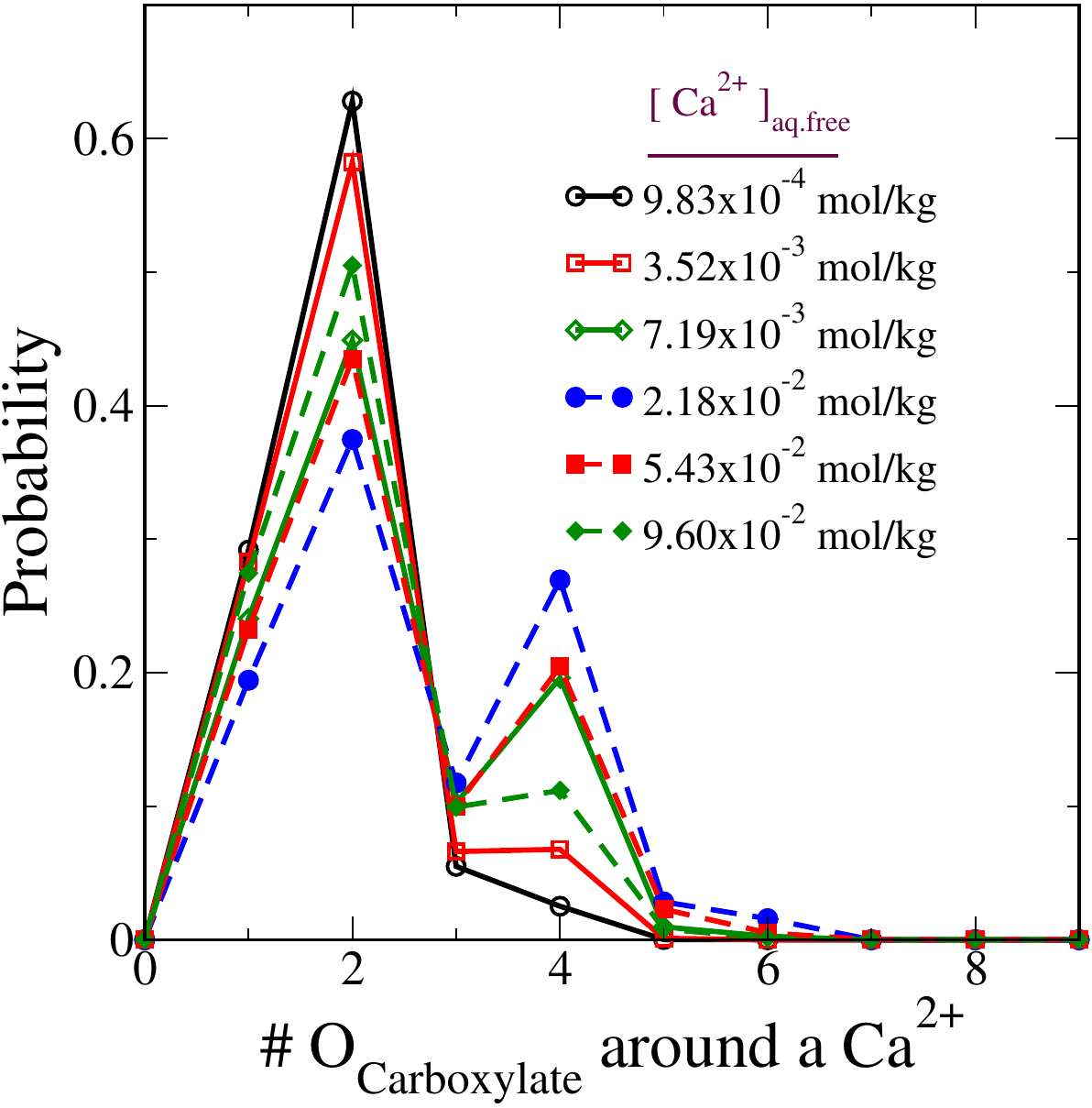}
    \caption{Probability of finding certain number of carboxylate oxygen atoms ($\mathrm{O_{Carboxylate}}$) on a 32-mer polyacrylate chain around a \ce{Ca^{2+}} ion.}
    \label{fig:BindEnv}
\end{figure}

The various binding modes and their coupling to polyelectrolyte conformation could impact the number of \ce{Ca^{2+}} ions sequestered. We determine a binding mode by tracking the number of \revDY{carboxylate oxygen atoms} that are around a \ce{Ca^{2+}} ion. From the radial distribution of \revDY{carboxylate oxygen atoms} around a \ce{Ca^{2+}} ion, we identify that their most probable separation is about 0.35 nm (see Supporting Information). We compute the probability of finding \revDY{a varying number of carboxylate oxygen atoms} within 0.35 nm from a \ce{Ca^{2+}} ion and report this as a function of $\mathrm{Ca^{2+}_{{aq.free}}}$ in \Cref{fig:BindEnv}.

We observe that the low binding mode, in which two carboxylate oxygen atoms bind to a given \ce{Ca^{2+}} ion, remains dominant at all concentrations of $\mathrm{Ca^{2+}_{{aq.free}}}$ studied. However, with an increase in $\mathrm{Ca^{2+}_{{aq.free}}}$, the high binding mode corresponding to ion-bridging becomes more favorable. Interestingly, when $\mathrm{Ca^{2+}_{{aq.free}}}$ exceeds $2.18 \times 10^{-2}$ mol/kg, this trend reverses: further increases in $\mathrm{Ca^{2+}_{{aq.free}}}$ promote the low binding mode once more. We hypothesize that this non-monotonic trend in the population of different binding modes arises from the enhanced electrostatic screening at higher \ce{Ca^{2+}} concentrations.

\begin{figure}[htbp]
    \centering
    \includegraphics[width=3 in]{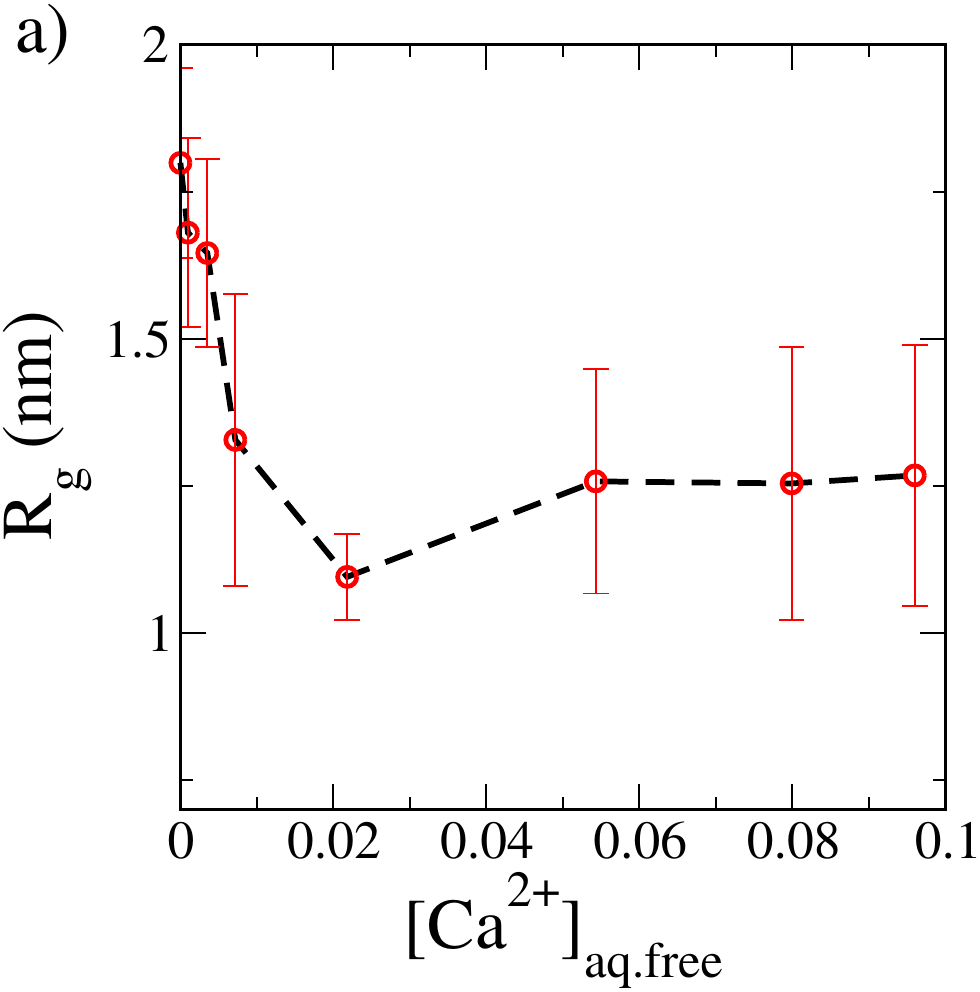}
    \vspace{10pt} 
    \includegraphics[width=3 in]{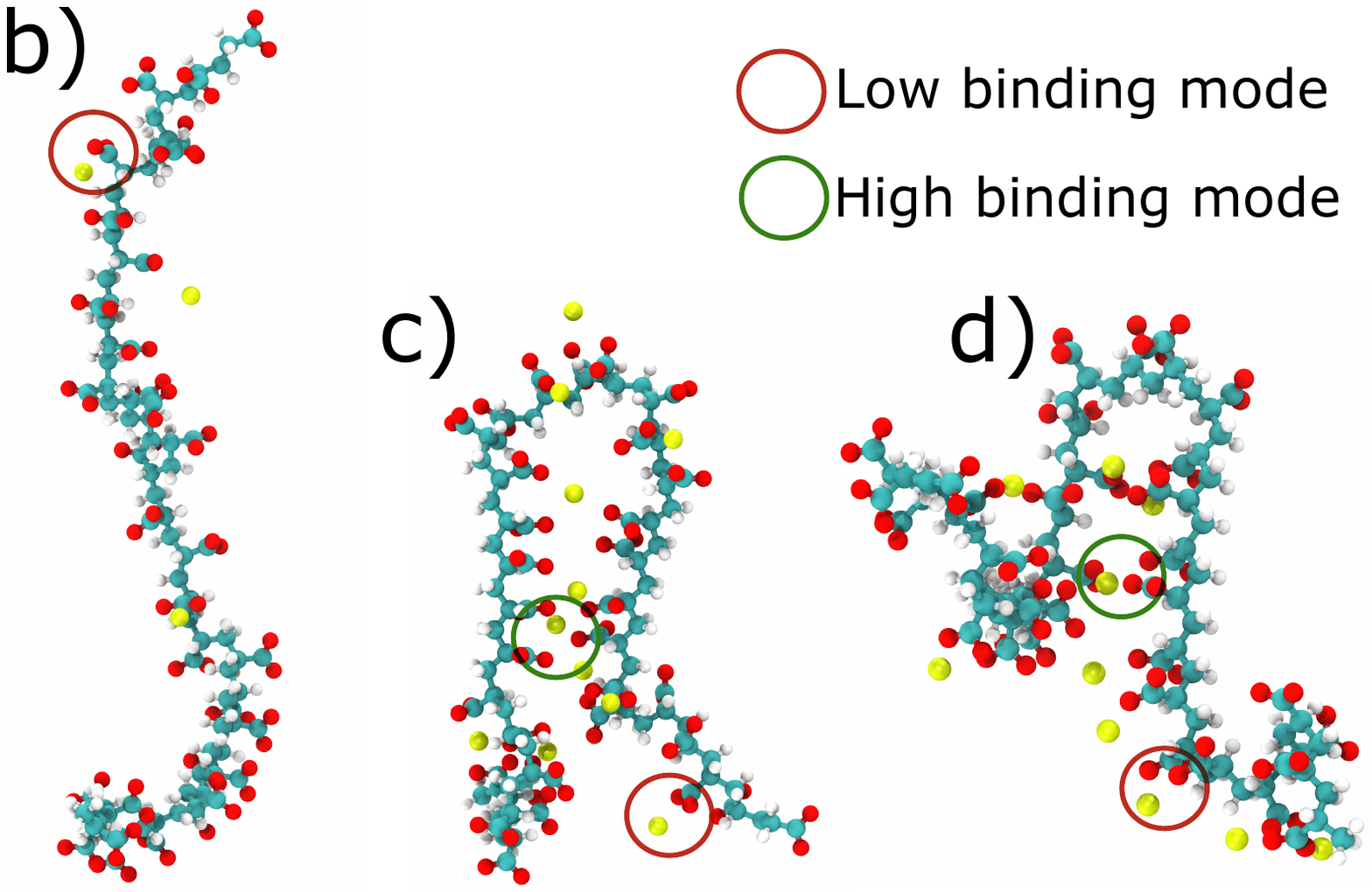}
    \caption{a) Radius of gyration of 32-mer polyacrylate chain at different concentrations of $\left[\mathrm{Ca^{2+}_{{aq.free}}}\right]$. Dominant binding modes at b) low, c) moderate, and d) high $\mathrm{\left[Ca^{2+}\right]_{aq.free}}$ concentrations. The conformation in (c) corresponds to the $\mathrm{R}_\mathrm{g}$ minimum in (a)}
    \label{fig:RgCa}
\end{figure}

Nevertheless, at moderate and high concentrations of $[\mathrm{Ca^{2+}_{{aq.free}}}$, nearly $\tfrac{1}{3}$ of the binding events are due to the high binding mode. This high coordination binding environment is seen to have a large impact on the polymer size and conformation. We investigate the coupling between the binding modes and the polyelectrolyte chain conformation by tracking the chain radius of gyration ($\mathrm{R_g}$) as a function of $\mathrm{Ca^{2+}_{{aq.free}}}$. We report these results in \Cref{fig:RgCa}.

We note from \Cref{fig:RgCa}(a) that at low concentrations of $\mathrm{Ca^{2+}_{{aq.free}}}$, where the population of the high binding mode is insignificant, the polymer chain adopts an extended conformation with an $\mathrm{R_g}$ of approximately 1.65--1.9 nm (\Cref{fig:RgCa}(b)). At these concentrations, \ce{Ca^{2+} ions} bind to at most one carboxylate group on the polyacrylate chain. As $\mathrm{Ca^{2+}_{{aq.free}}}$ increases and the population of high coordination binding sites subsequently increases, the polymer conformation transitions to a collapsed state (\Cref{fig:RgCa}(c)). Here, the high coordination binding sites, which bridge two strands of the polyelectrolyte chain, are nearly as prominent as the low coordination binding sites located on the solvent-exposed side of each strand. Intriguingly, in solutions with high concentrations of \ce{Ca^{2+}} ions, although some of the bridging events are disrupted, the conformation of a polyelectrolyte chain still resembles that of a collapsed state (\Cref{fig:RgCa}(d)). The disruption of the bridging events due to enhanced screening from the other ions in the system increases the conformational flexibility and hence the average chain size.

\begin{figure}[htbp]
    \centering
    \includegraphics[width=3 in]{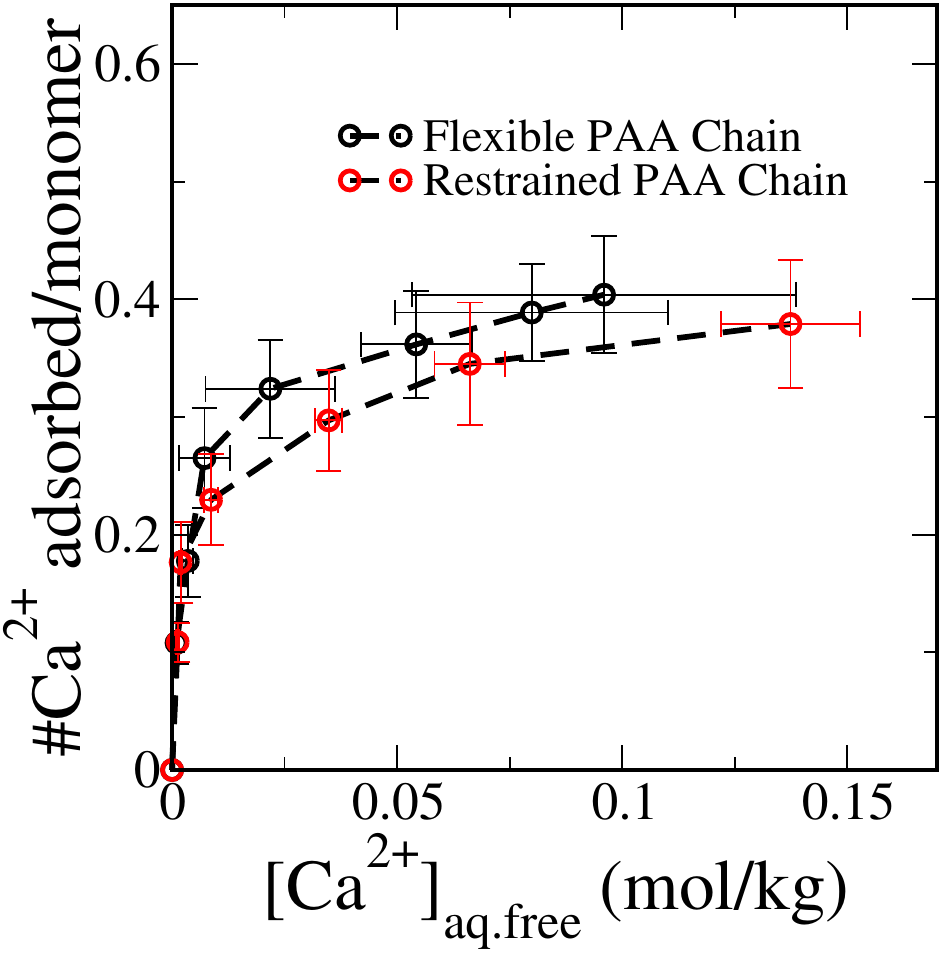}
    \caption{\ce{Ca^{2+}} ion binding isotherm to a polyacrylate chain with 32 \revDY{repeat units} restrained to an extended coil conformation and unrestricted.}
    \label{fig:Restr}
\end{figure}

Since the high coordination binding sites are energetically more favorable, we anticipate that access to a larger population of these binding sites would enhance the ability of a polyelectrolyte chain to sequester \ce{Ca^{2+}} ions. To explore this, we investigate the inverse problem; we limit the polyelectrolyte chain to only binding sites with low coordination environments and construct the adsorption isotherm. We achieve this by first identifying the polymer conformation that corresponds to the most probable $\mathrm{R}_\mathrm{g}$ in a system with no added \ce{Ca^{2+}} ions. Utilizing this chain conformation, with harmonic restraints imposed, we prepare polyelectrolyte solutions with added \ce{Ca^{2+}} ions. We compute the \ce{Ca^{2+}} adsorption isotherm from these systems and compare it to that obtained from unrestrained simulations reported earlier. We show these results in \Cref{fig:Restr}.

We find that at low concentrations, the number of \ce{Ca^{2+}} ions adsorbed on the polyacrylate chain in both the restrained and unrestrained simulations is indistinguishable. This is because, at lower concentrations of \ce{Ca^{2+}} ions in the solution, unrestrained polyelectrolyte chains can only access the low coordination binding sites. However, we see a noticeable difference at moderate and high concentrations, where the unrestrained polyacrylate chain favors the high coordination binding sites. The restrained simulations, which do not have access to high binding modes, consistently adsorb fewer \ce{Ca^{2+}} ions per chain when compared to the unrestrained simulations.

\revIP{These observations indicate that a polyelectrolyte chain's ability to efficiently chelate} \ce{Ca^{2+}} \revIP{ions is impacted by its access to a large number of high coordination binding sites. These sites are energetically more favorable for ion binding, and as such, they provide a more stable environment for the ions, leading to a more effective sequestration. This insight could prove useful in applications where selective and efficient ion capture is paramount, such as in water treatment processes or biomedical applications.}


\section{Summary}
Using all-atom molecular simulations coupled with a free energy perturbation approach, we constructed an adsorption isotherm to describe the binding of \ce{Ca^{2+}} ions to a model polyacrylate chain. Analysis of the adsorption isotherm revealed that the per-monomer \ce{Ca^{2+}} ion binding capacity of a fully charged polyelectrolyte chain saturates at a value of $0.40\pm0.05$. This saturation value correlates with the binding modes accessible to \ce{Ca^{2+}} ions as they bind to the polyelectrolyte chain. Two predominant binding modes were identified: one mode involves \ce{Ca^{2+}} ions binding to at most two carboxylate oxygen atoms on a polyacrylate chain, and the other involves \ce{Ca^{2+}} ions binding to four or more carboxylate oxygen atoms.

The population of low binding mode sites remains high across all concentrations of \ce{Ca^{2+}} in the solution. Nevertheless, at least one-third of the binding events at moderate and high concentrations of \ce{Ca^{2+}} ions in the solution are defined by the high binding mode events. These binding events, while responsible for enhancing the \ce{Ca^{2+}} ion binding capacity of a polyelectrolyte chain, also lead to the collapse of the polyelectrolyte chain's conformation. The solution concentration of \ce{Ca^{2+}} ions corresponding to the conformational transition falls within the range close to the supernatant (polymer-poor) side of the phase behavior reported by Sabbagh et al.\cite{Sabbagh2000} The adsorption isotherm constructed in this study suggests that, at these concentrations, the collapsed polyelectrolyte chain had attained the saturation value of the \ce{Ca^{2+}} binding capacity. 

The collapsed polyelectrolyte chain, with all available binding sites saturated, may indicate the onset of putative phase separation. This observation carries important implications for the chelating capacity of a polyelectrolyte towards \ce{Ca^{2+}} ions, and consequently, for its antiscalant activity.  Yet, polyelectrolyte-divalent ion complex falling out of equilibrium is inherently a multi-chain problem. In our follow up work,\cite{Glisman2023} we tackle this problem and  establish the conditions under which a polyelectrolyte in divalent salt solution falls out of equilibrium.
\label{Sec:Concl}




\begin{acknowledgement}
    This work was supported by the Dow Chemical Company through a University Partnership Initiative (UPI) grant. We benefited greatly from the discussions with our Dow collaborators: Thomas Kalantar, Christopher Tucker, Larisa Reyes, and Meng Jing. S.M. thanks Dr. Chris Balzer for discussions and useful comments.
\end{acknowledgement}

\begin{suppinfo}

In the supporting document, we first demonstrate overcharging of the polyacrylate-\ce{Ca^{2+}} complex when using non-polarizable force fields with full charges on ions. Then, we validate scaled charge force fields for modeling polyacrylate conformations. Following that, we present evidence for meaningful sampling of polyelectrolyte-\ce{Ca^{2+}} complexes in an aqueous solution using a non-polarizable force field with electronic continuum correction and Hamiltonian Replica-Exchange Molecular simulations. Finally, we discuss how we established chemical potential equivalence conditions and determined the concentration of free \ce{Ca^{2+}} ions in the solution that is in equilibrium with the system containing the polyacrylate chain.

\end{suppinfo}

\bibliography{CaAdsLit}

\end{document}


\tableofcontents
\section{Overcharging of Polyelectrolyte—ion Complex in Models with Full Ion Charges}

\begin{figure}[htbp]
  \centering
  \begin{subfigure}{0.5\textwidth}
    \centering
    \includegraphics[width=2.5 in]{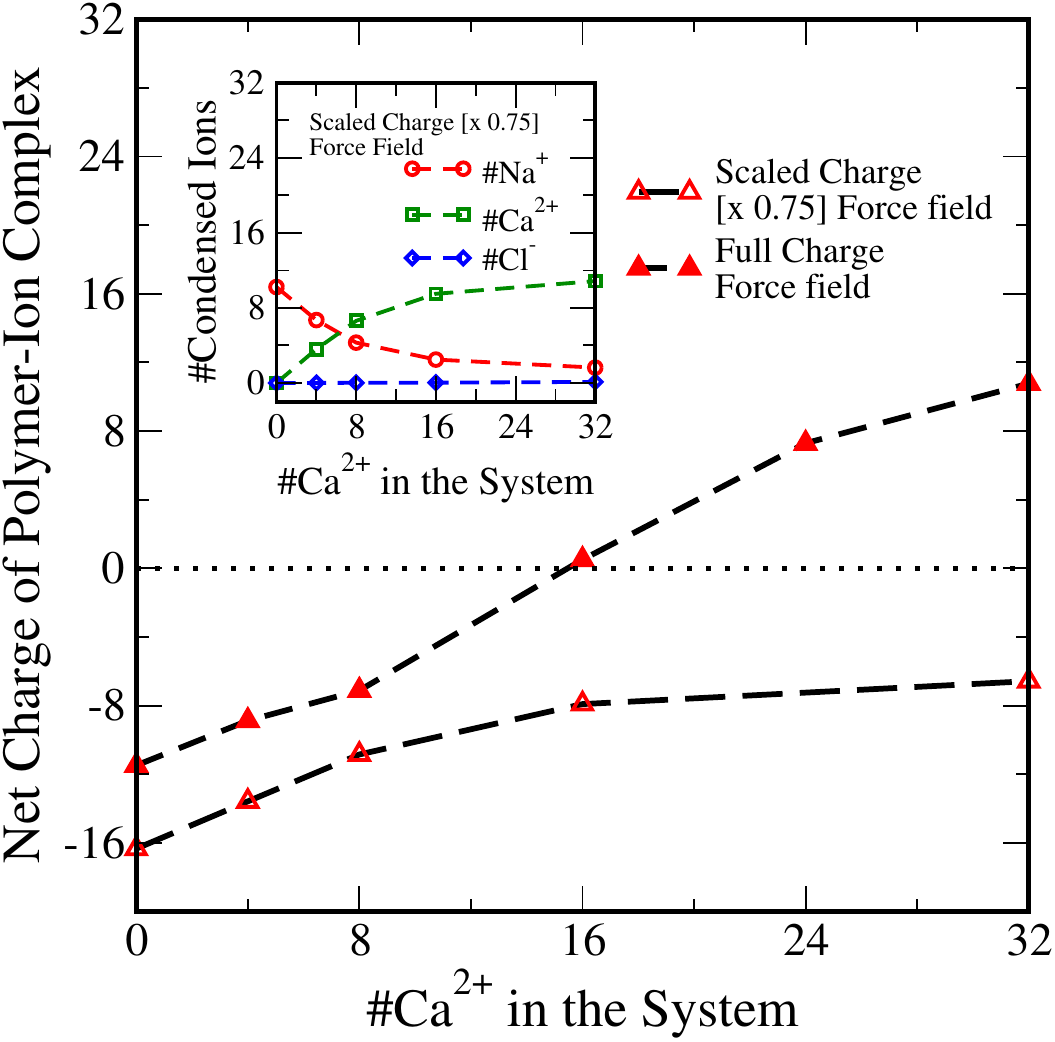}
  \end{subfigure}%
  \begin{subfigure}{0.5\textwidth}
    \centering
    \includegraphics[width=2.75 in]{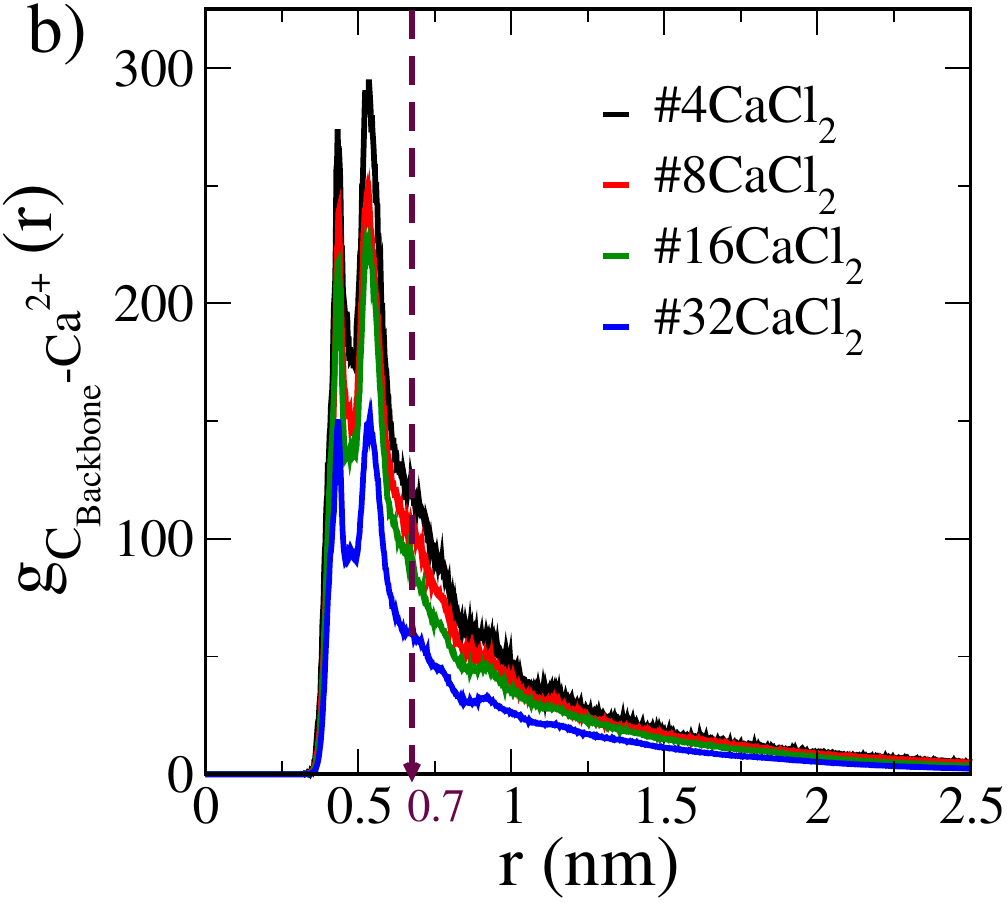}
  \end{subfigure}
  \caption{a) Net Charge of a Polyelectrolyte-Condensed Ion Complex: A Comparison between Results from a Full Charge and Scaled Charge Force Field. b) Radial Distribution of \ce{Ca^{2+}} Ions around a Backbone Carbon on the Polyacrylate Chain Computed from Simulations Using a Scaled Charge Force Field. }
  \label{sfig:NetChrg}
\end{figure}

Molecular dynamics simulations, which utilize non-polarizable force fields with "full charge" models for ions, tend to overestimate the binding of dissolved ions to the polyelectrolyte chain. We demonstrate this in \Cref{sfig:NetChrg}(a) for a fully charged polyacrylate with 32 monomers per chain and sodium ions added to balance the charge on the polymer, in an aqueous \ce{CaCl2} solution. In the figure, we report the net charge of the polyacrylate-ion complex as a function of the number of \ce{Ca^{2+}} ions in the system. From the radial distribution of \ce{Ca^{2+}} ions around a backbone carbon on the polyacrylate chain (see \Cref{sfig:NetChrg}(b)), we identify that the \ce{Ca^{2+}} ions are most likely to be found at a separation of 0.7 nm from the polymer carbon. We identify all the ions (\ce{Ca^{2+}}, \ce{Na^{+}}, \ce{Cl^{-}}) that are located within 0.7 nm from the polymer backbone and label them as condensed ions. The net charge of the polyacrylate-ion complex is calculated by summing the partial charges of the condensed ions.

We observe from \Cref{sfig:NetChrg}(a) that simulations with the "full charge" model yield a shift in the net charge of the polyacrylate-ion complex from negative to positive with an increase in \ce{Ca^{2+}} concentration in the system. In contrast, the "scaled charge" models described in Section 2 predict an increase in the net charge, which eventually plateaus at a negative value for higher \ce{Ca^{2+}} concentrations. This observation from the "scaled charge" model aligns with recent potentiometric titrations conducted by Gindele et al. \cite{Gindele2022}.

\section{Validating Scaled Charge Forcefield for modeling Polyelectrolyte Conformations}

\begin{figure}[htbp]
    \centering
    \includegraphics[width=2.5 in]{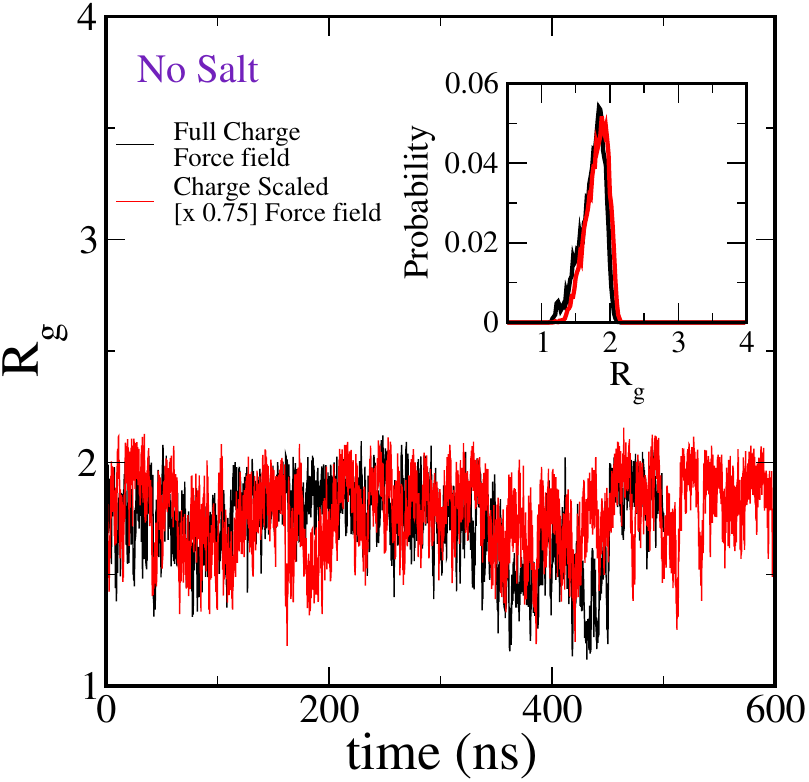}
    \caption{Temporal variation of the radius of gyration($\mathrm{R_g}$ of a polyacrylate chain in an aqueous solution without any added salt}
    \label{sfig:RgNoSalt}
\end{figure}

It is not clear beforehand whether the scaled charge-corrected polyacrylate chain spans the desired conformational space. Mintis et al. \cite{Mintis2019} rigorously tested their full charge models against experimentally determined properties of polyacrylate chains in salt-free solutions. We hypothesize that if the scaled charge model for the polyacrylate chain spans the same conformational space, then it will not significantly alter their structural properties. We confirm this observation through the data presented in Figure \ref{sfig:RgNoSalt}.

In the figure, we present the temporal variation of the radius of gyration ($\mathrm{R_g}$) of the polyacrylate chain in a salt-free aqueous solution. We observe that the range of $\mathrm{R_g}$ values covered by both the full charge models and the scaled charge models are very similar. This observation is further supported by the $\mathrm{R_g}$ probability distribution shown in the inset of the figure.

\section{Meaningful Sampling with Electronic Continuum Correction and Hamiltonian Replica-Exchange Molecular Simulations}
\begin{figure}[htbp]
  \centering
  \begin{subfigure}{0.5\textwidth}
    \centering
    \includegraphics[width=2.5 in]{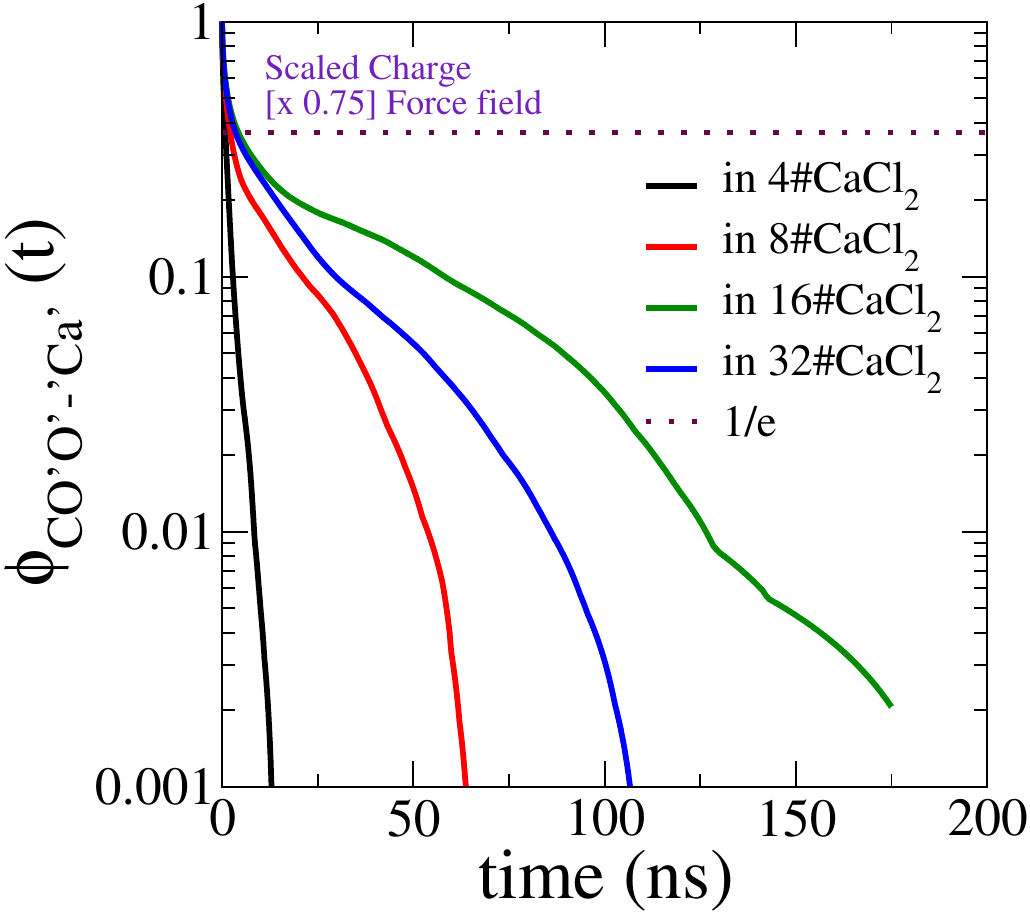}
  \end{subfigure}%
  \begin{subfigure}{0.5\textwidth}
    \centering
    \includegraphics[width=2.75 in]{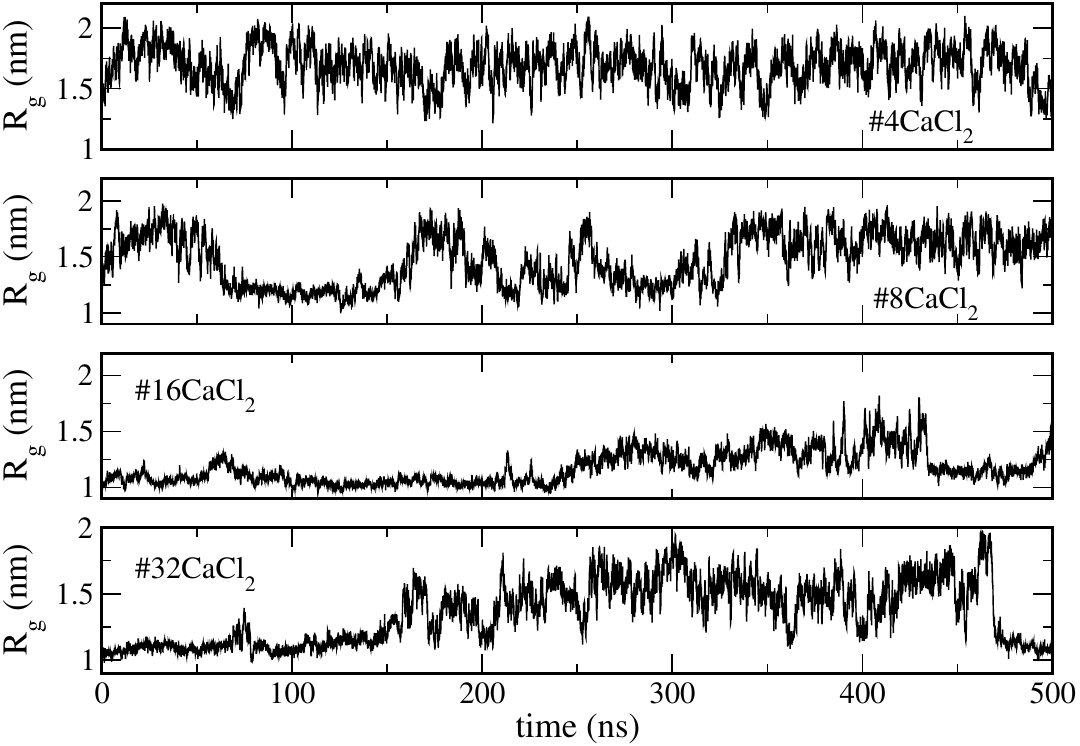}
  \end{subfigure}
  \caption{a) Binding/unbinding relaxation of a carboxylate oxygen --\ce{Ca^{2+}} ion pair with the scaled charge force field. The function $\phi\left(t\right)$ represents the autocorrelation function depicting ion-pair survival probability\cite{Impey1983}. b) Temporal variation of the radius of gyration ($\mathrm{R_g}$) of a polyacrylate chain in an aqueous solution with varying concentrations of \ce{CaCl2}. }
  \label{sfig:IPRg}
\end{figure}

Charge scaling has significantly improved the binding/unbinding relaxation times of polyacrylate-\ce{Ca^{2+}} interactions. \Cref{sfig:IPRg}(a) illustrates the decay of the ion-pair survival probability autocorrelation function\cite{Impey1983} to zero within the time scales achievable in molecular simulations using modern GPU architecture. However, a trajectory spanning 500 ns proved inadequate for sampling the conformational space with statistical certainty. The temporal variation of $\mathrm{R_g}$, as reported in \Cref{sfig:IPRg}(b), indicates that the conformational space explored by polyelectrolyte chains in an aqueous solution with a concentration equal to or greater than 8\ce{CaCl2} is not representative of an equilibrium distribution. 

\begin{figure}[htbp]
    \centering
    \includegraphics[width=2.5 in]{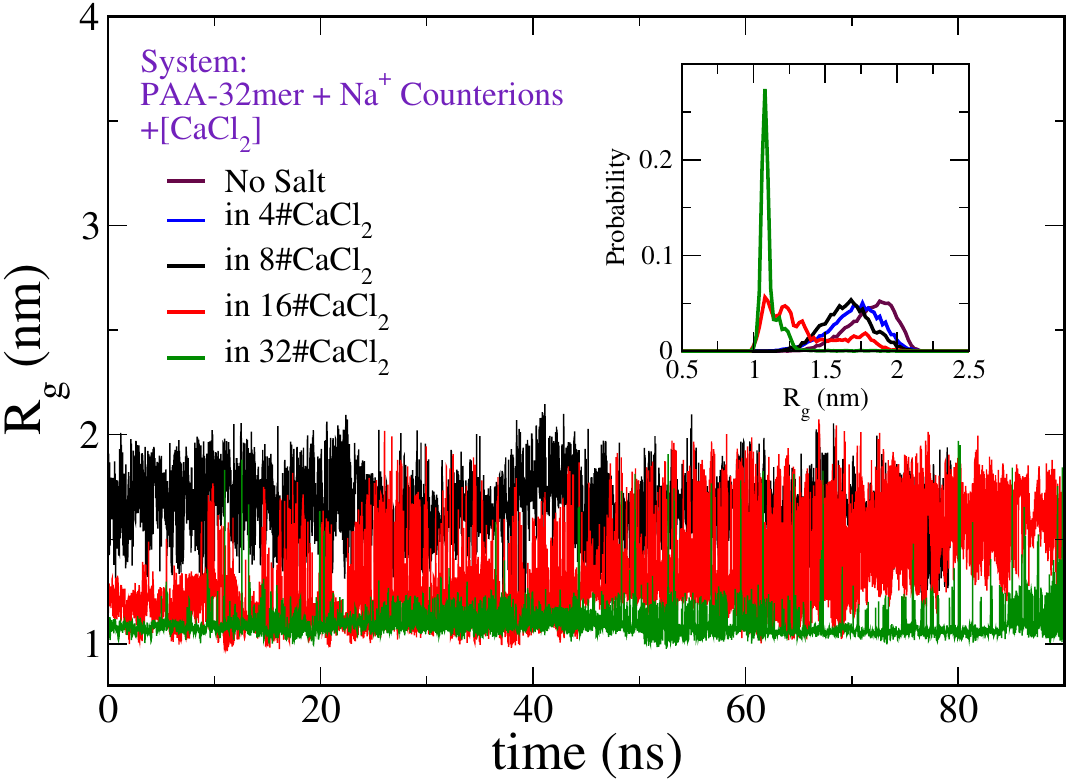}
    \caption{Predictions for the radius of gyration ($\mathrm{R_g}$) of a polyacrylate chain in an aqueous solution with varying concentrations of \ce{CaCl2}, computed using Hamiltonian Replica Exchange simulations with a scaled charge force field.}
    \label{sfig:RgHREMD}
\end{figure}

Estimating the required length of a simulation trajectory to generate an equilibrium distribution is challenging. To overcome this issue, we use a Hamiltonian replica Exchange protocol described in Section 2 that specifically biases polyacrylate-Ion interactions in the system.  In \Cref{sfig:RgHREMD}, we present the temporal evolution of the radius of gyration of a fully charged PAA-32mer chain at different \ce{Ca^{2+}} concentrations in the system, computed from the HREMD simulation. A 100 ns trajectory already demonstrates that the simulations have explored many possible chain conformations under given system conditions.

\section{Determining $\left[\ce{Ca^{2+}}\right]_\mathrm{aq.free}$ from the equivalence of $\mu^\mathrm{system}_{\ce{CaCl2}}$ and $\mu^\mathrm{solution}_{\ce{CaCl2}}$}

\begin{figure}[htbp]
    \centering
    \includegraphics[width=2.5 in]{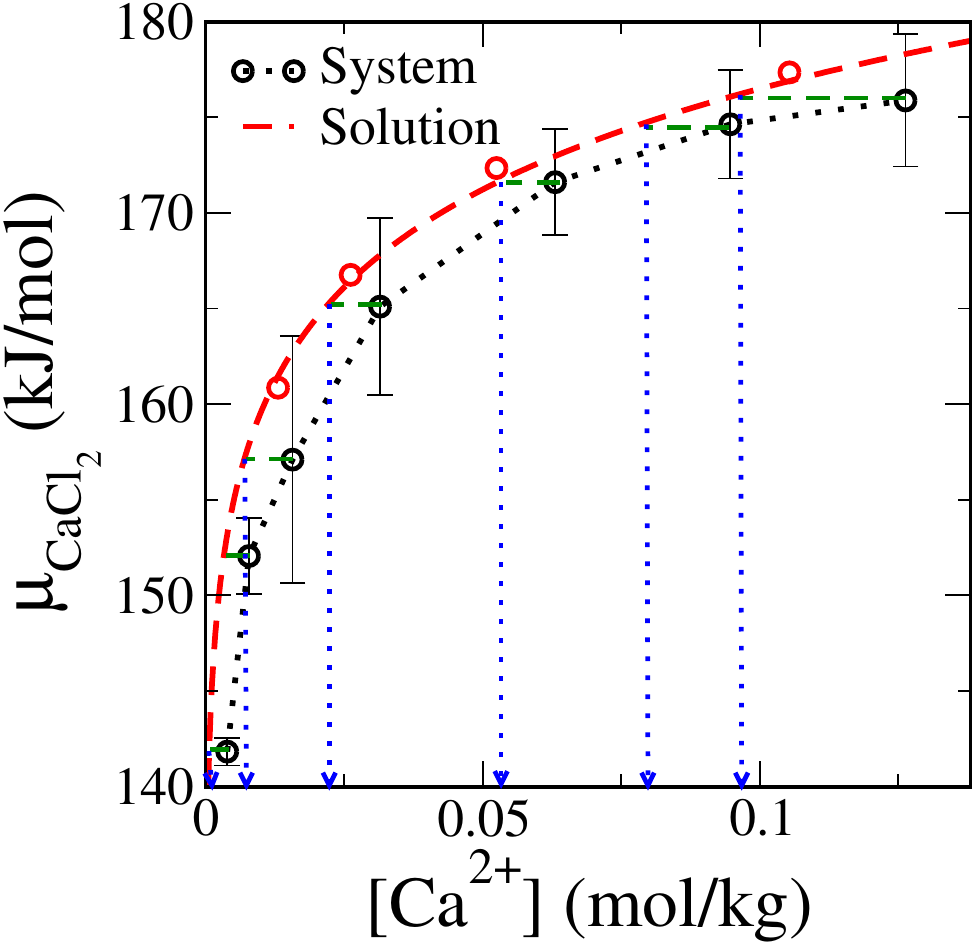}
    \caption{The chemical potential of \ce{CaCl2} is computed from both the system with the polyacrylate chain and the solution without the polyacrylate chain. The horizontal green dashed lines indicate the equivalence of the \ce{CaCl2} chemical potential between the System and the Solution. The vertical blue dotted lines indicate the concentration of free calcium ions in the solution that are in equilibrium with the system of interest.}
    \label{sfig:Chempot}
\end{figure}

$\left[\ce{Ca^{2+}}\right]_\mathrm{aq.free}$ represents the concentration of free \ce{Ca^{2+}} ions in the solution that are in equilibrium with those adsorbed on a polyacrylate chain. We determined $\left[\ce{Ca^{2+}}\right]_\mathrm{aq.free}$ by equating the chemical potential of \ce{CaCl2} in the system with the polyacrylate chain ($\mu^\mathrm{system}_{\ce{CaCl2}}$) to the chemical potential without the polyacrylate chain ($\mu^\mathrm{solution}_{\ce{CaCl2}}$). We report these chemical potentials in \Cref{sfig:Chempot} and establish their equivalence conditions.

We note that the lowest $\left[\ce{CaCl2}\right]_{aq}$, without polyacrylate chain, we simulated in this work is 0.013 mol/kg. As practised by Pangiotopolous et al\cite{Young2018}, we assumed this concentration of $\left[\ce{CaCl2}\right]_\mathrm{aq} = 0.013 \mathrm{mol/kg}$ to be sufficiently low for Debye-Huckle limiting law for electrolyte solutions to hold true. For any concentrations of $\left[\ce{CaCl2}\right]_\mathrm{aq} < 0.013 \mathrm{mol/kg}$, we used Debye-Huckle limiting law to compute $\mu^\mathrm{solution}_{\ce{CaCl2}}$.
\bibliography{CaAdsLit}